\magnification\magstephalf
\overfullrule 0pt

\font\rfont=cmr10 at 10 true pt
\def\ref#1{$^{\hbox{\rfont {[#1]}}}$}


\font\fourteenbf=cmbx12 scaled\magstep1
\font\fourteenbfit=cmbxti10 scaled\magstep2


  \def\la{\lambda}
 \def\L {\Lambda} \def\G {\Gamma}

\def\pmb#1{\setbox0=\hbox{#1}
 \kern.05em\copy0\kern-\wd0 \kern-.025em\raise.0433em\box0 }

\def\slash{/\kern-.5em}

\def \half {{\scriptstyle {1 \over 2}}}

 %


\def\boxit#1{\vbox{\hrule\hbox{\vrule\kern1pt\vbox
{\kern1pt#1\kern1pt}\kern1pt\vrule}\hrule}}

\def\h{\hfill\break}
\parskip=6pt
\parindent=0pt
\hsize=17truecm\hoffset=-5truemm
\vsize=24truecm
\def\footnoterule{\kern-3pt
\hrule width 17truecm \kern 2.6pt}


\catcode`\@=11 

\def\nolabels{\def\wrlabeL##1{}\def\eqlabeL##1{}\def\reflabeL##1{}}
\def\writelabels{\def\wrlabeL##1{\leavevmode\vadjust{\rlap{\smash%
{\line{{\escapechar=` \hfill\rlap{\sevenrm\hskip.03in\string##1}}}}}}}%
\def\eqlabeL##1{{\escapechar-1\rlap{\sevenrm\hskip.05in\string##1}}}%
\def\reflabeL##1{\noexpand\llap{\noexpand\sevenrm\string\string\string##1}}}
\nolabels
\global\newcount\refno \global\refno=1
\newwrite\rfile
\def\defref{$^{{\hbox{\rfont [\the\refno]}}}$\nref}
\def\nref#1{\xdef#1{\the\refno}\writedef{#1\leftbracket#1}%
\ifnum\refno=1\immediate\openout\rfile=refs.tmp\fi
\global\advance\refno by1\chardef\wfile=\rfile\immediate
\write\rfile{\noexpand\item{#1\ }\reflabeL{#1\hskip.31in}\pctsign}\findarg}
\def\findarg#1#{\begingroup\obeylines\newlinechar=`\^^M\pass@rg}
{\obeylines\gdef\pass@rg#1{\writ@line\relax #1^^M\hbox{}^^M}%
\gdef\writ@line#1^^M{\expandafter\toks0\expandafter{\striprel@x #1}%
\edef\next{\the\toks0}\ifx\next\em@rk\let\next=\endgroup\else\ifx\next\empty%
\else\immediate\write\wfile{\the\toks0}\fi\let\next=\writ@line\fi\next\relax}}
\def\striprel@x#1{} \def\em@rk{\hbox{}} 
\def\lref{\begingroup\obeylines\lr@f}
\def\lr@f#1#2{\gdef#1{\defref#1{#2}}\endgroup\unskip}
\newwrite\lfile
{\escapechar-1\xdef\pctsign{\string\%}\xdef\leftbracket{\string\{}
\xdef\rightbracket{\string\}}}

\def\writestop{\def\writestoppt{\immediate\write\lfile{\string\p
ageno%
\the\pageno\string\startrefs\leftbracket\the\refno\rightbracket%
\string\def\string\secsym\leftbracket\secsym\rightbracket%
\string\secno\the\secno\string\meqno\the\meqno}\immediate\closeout\lfile}}
\def\writestoppt{}\def\writedef#1{}
\catcode`\@=12 
\input epsf.tex
\rightline{DAMTP 97/118}
\rightline{M/C-TH 97/14}
\bigskip
\centerline{\fourteenbf FINAL STATE INTERACTIONS IN {\fourteenbfit WW} 
PRODUCTION}

\vskip 10mm
\centerline{A Donnachie}
\vskip 3mm
\centerline{Department of Physics and Astronomy, University of Manchester,
Manchester M13 9PL, UK}
\vskip 6mm
\centerline{P V Landshoff}
\vskip 3mm
\centerline{DAMTP, University of Cambridge, Cambridge CB3 9EW, UK}
\vskip 10mm
{\bf{Abstract}}

It is shown that colour transparency causes nonperturbative colour-singlet
final-state interactions to have a negligible effect on the production rate
and the dijet mass spectra in $e^+e^-\to WW$.
However, the same cannot be said of nonperturbative colour-octet exchange,
for which we show that there are
indications of observable effects, 
though we are unable to present precise estimates.
\vskip 2cm

{\bf Introduction}

An important aim at LEP2 is to measure  the mass of the 
$W$-boson to high accuracy, perhaps to within 50 MeV\defref\mass{
{\sl Determination of the mass of the W-boson},
Report of the Workshop on Physics at LEP2, CERN Yellow report CERN-96-01 
(hep-ph/9602352)
}. A preferred method, because it offers the prospect of the most
statistics, is to produce a pair of $W$'s, each of which decays into
a pair of quark jets. The invariant mass of each jet pair is then measured.

An obvious question is whether final-state interactions among the jets
will cause a problem: the $W$-particles have a short lifetime, so the quarks
are close together when they are produced and interactions among them could well
be significant. There have been several calculations of colour exchange 
among the quarks\defref\ellis{
J Ellis and K Geiger, Physical Review D54 (1996) 1967\h
G Gustafson and J H\"akkinen, Z Physik C64 (1994) 659\h
G Gustafson, U Petersson and P Zerwas, Physics Letters B209 (1988) 90
}\defref\sk{
T Sj\"ostrand and V Khoze, Z Physik C62 (1994) 281
},
mostly reaching the conclusion that the effect is quite small.

In this paper, we first examine the effect of colour-singlet exchange.
Because the relative energies of the quarks are large, we use 
soft-pomeron phenomenology\defref\zuoz{
P V Landshoff, Proc PSI school at zuoz (Villigen, 1994) hep-ph/9410250\h
J R Forshaw and D A Ross, {\sl Quantum chromodynamics and the pomeron},
Cambridge University Press (1997)
} to model this exchange. While this phenomenology is well-established and
has had many successes, its extension to the present problem does involve
uncertainty. Nevertheless, our conclusion that there is a negligibly
small effect is probably reliable.
This result comes about because colour-transparency effects\defref\bertsch
{G Bertsch, S J Brodsky, A S Goldhaber and J F Gunion, Physical Review Letters
47 (1981) 297
} 
suppress colour-singlet exchange.

We then go on to perform a similar calculation of colour-octet exchange.
We model this exchange using Cornwall's solution to the Schwinger-Dyson
equations\defref\cornwall{
J M Cornwall, Physical Review D26 (1982) 1453
}, which provides a well-motivated way to handle the nonperturbative region
of gluon exchange.
In the case of perturbative gluon exchange, there is an infrared divergence,
corresponding to soft exchange,
which is cancelled by adding in the contribution from soft gluon emission.
However, these infrared divergences are not genuine, because
the nonperturbative corrections to the propagator, which have their origin
in confinement, remove them. The Cornwall formalism
does not provide a method of calculating the gluon emission, but since
it effectively gives the gluon a mass, any emission that does occur cannot be
soft. Hence there is nothing to cancel the correction to the cross section
from soft gluon exchange. We therefore calculate soft nonperturbative gluon
exchange and maintain that it is likely to provide a lower bound to the
true correction to the cross section from colour-octet exchange.
We find that this lower bound is far from small.
\bigskip
{\bf Formalism}

We consider the differential cross-section ${d^2\sigma/dM_1\, dM_2}$ for the 
$e^+e^-$  annihilation into $WW$ and the decay of each $W$ into a 
quark-antiquark pair, $WW\to q\bar q q\bar q$,
with $M_1$ and $M_2$ the invariant masses of the jet pairs.
One of the Born diagrams for this process is shown in figure 1a. We concentrate 
on the corrections to this diagram; we do not expect any significant difference
in the result for the other Born diagram, in which the $e^+e^-$ transform
into $WW$ through neutrino exchange.

We suppose that the interactions among the final quarks are two-body
interactions, as depicted in figure 1b. We suppose further that they  
do not flip the helicities of the quarks.
Then the ratio of the amplitude of 
figure 1b to that of figure 1a is independent of the spin states of the
leptons and the quarks and is
$$
R(P_1,P_2,p_1,p_2)=i{(P_1^2-M_W^2)(P_2^2-M_W^2)\over (2\pi)^4}
\int d^4\Delta {T_{qq}(p_1,p_2,\Delta )\over \big ((P_1+\Delta )^2-M_W^2 \big )
\big ((P_2-\Delta )^2-M_W^2 \big )}
\eqno(1)
$$
Here the $W$ mass is complex, $M_W=M-i\Gamma$.
The quark-quark scattering amplitude $T_{qq}$ includes the two
off-shell lines $\bar p_1=p_1+\Delta$ and 
$\bar p_2=p_2-\Delta$. It is averaged over the quark
spin states. 
We have to sum (1) over the four different ways of attaching the interaction to
the quark and antiquark lines.

\topinsert
\centerline{\hfill{\hbox{\epsfxsize=60truemm\epsfbox{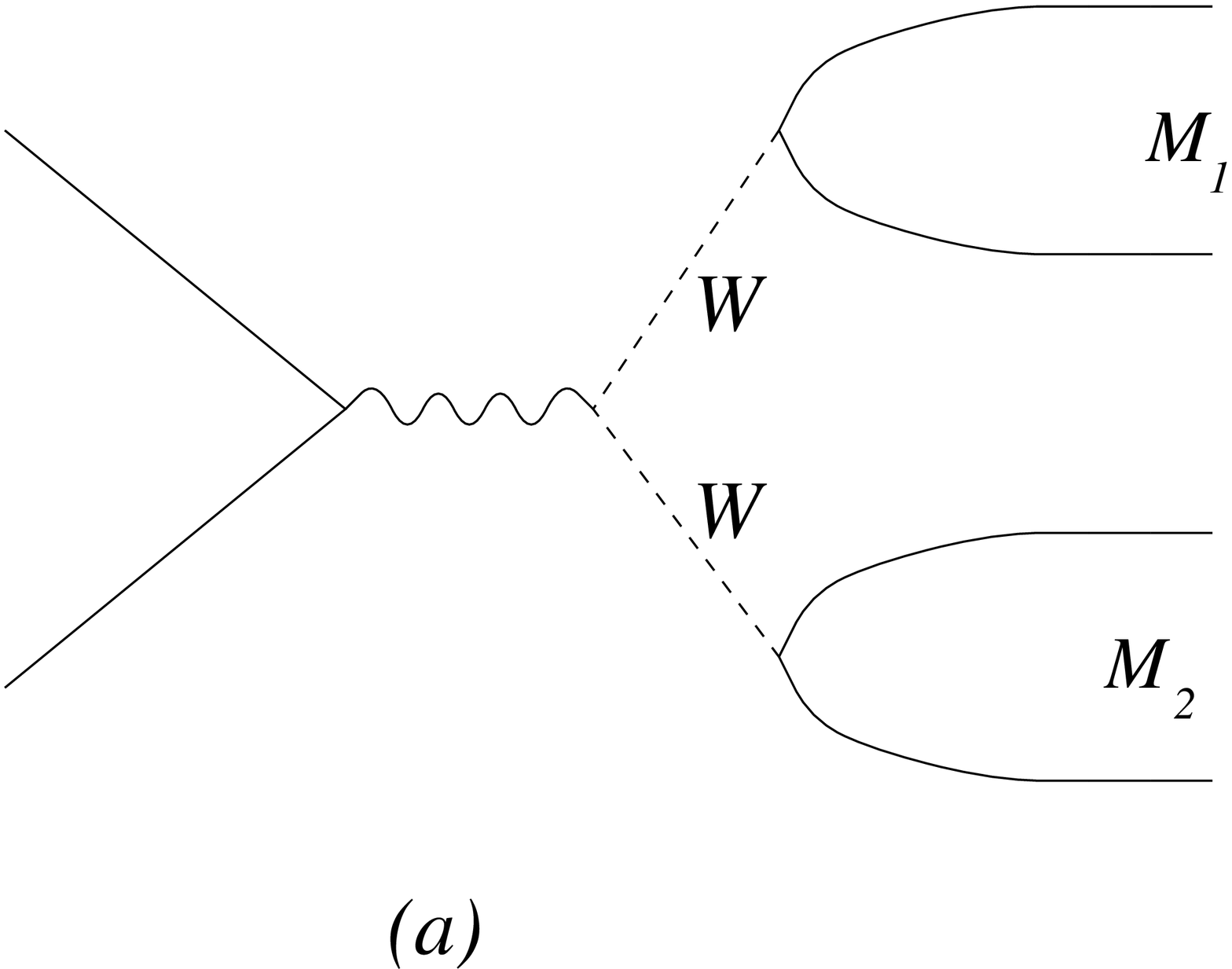}}}
\hfill{\hbox{\epsfxsize=80truemm\epsfbox{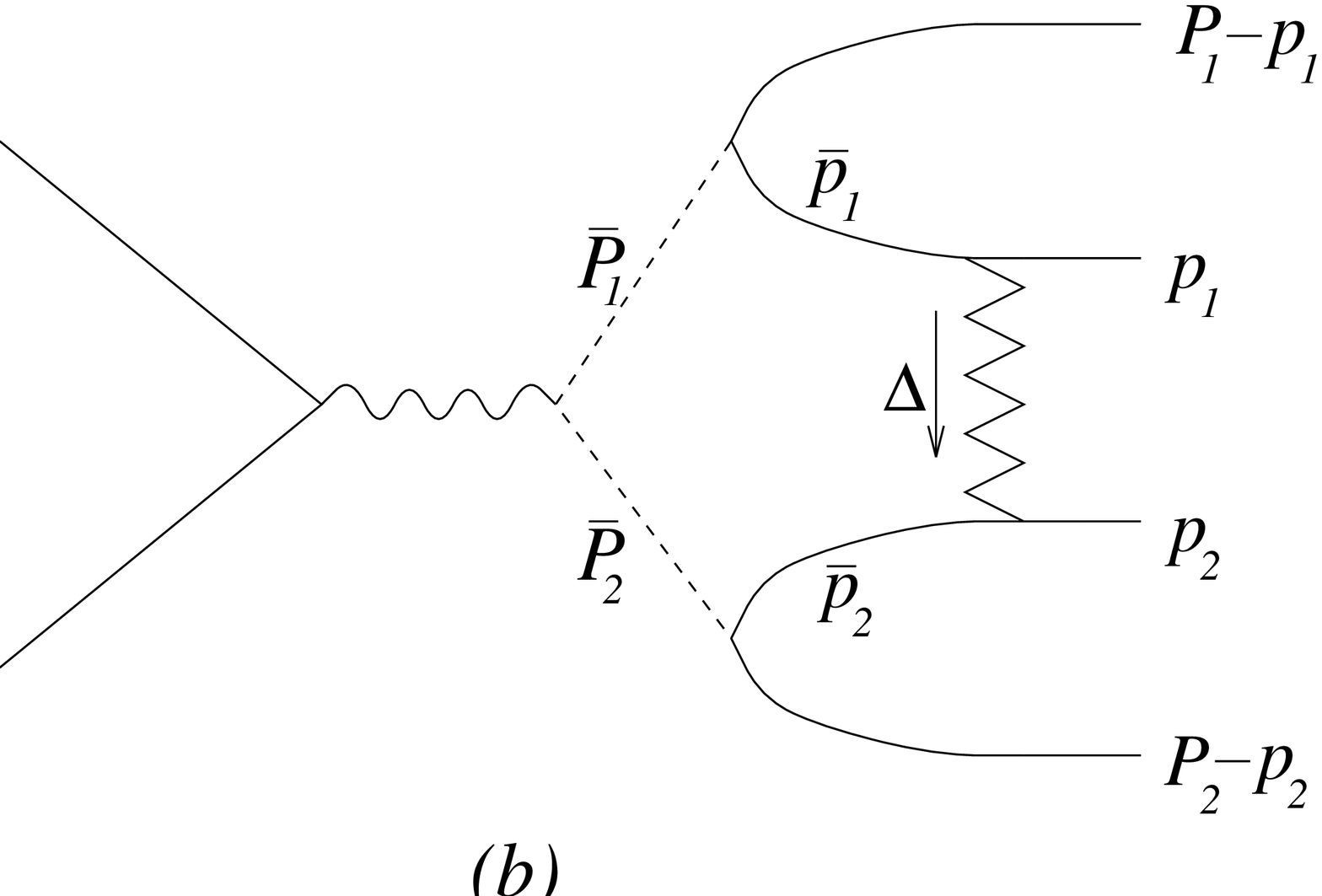}}}\hfill}
Figure 1: $e^+e^-\to $ 4 quark jets via $WW$, (a) the Born diagram and (b)
with final-state interaction.
\endinsert

We define $\nu = P_1\cdot P_2$.
It is useful to introduce linear combinations $R_1$ and $R_2$ of $P_1$ and
$P_2$ satisfying $R_1^2=0=R_2^2$:
$$
P_1=R_1+\lambda _1R_2 ~~~~~~~~~~~~~~~~~P_2=R_2+\lambda _2R_1
\eqno(2a)
$$
with
$$
\lambda _1={\nu - \sqrt{\nu ^2-M_1^2M_2^2}\over M_2^2}~~~~~~~~~~
\lambda _2={\nu - \sqrt{\nu ^2-M_1^2M_2^2}\over M_1^2}
\eqno(2b)
$$
Then $\nu _R=R_1\cdot R_2$ satisfies $\nu _R=\nu (1+\lambda _1\lambda _2)$.
We parametrise
$$
\Delta={\alpha\over 2\nu_R}R_1-{\beta\over 2\nu_R}R_2+\delta
\eqno(3a)
$$
where $\delta\cdot R_1=0=\delta\cdot R_2$, so that $\delta$ is a
two-dimensional anti-euclidean vector.  Then
$$
\int d^4\Delta={1\over 4\nu _R}\int d\alpha d\beta\, d^2\delta
\eqno(3b)
$$
We find that it is consistent to
assume that most of the contribution to the $\Delta$
integration will arise from values of $\alpha$ and $\beta$ that
are much less than $\nu _R$, so that 
$$
\hat t=\Delta ^2\sim\delta ^2
\eqno(4a)
$$
and the squared 4-momenta of the $W$'s are
$$\eqalignno{
\bar P_1^2=(P_1+\Delta)^2)&\sim A_1 -\beta +\lambda _1\alpha\cr
\bar P_2^2=(P_2-\Delta ^2)&\sim A_2 -\alpha +\lambda _2\beta
&(4b)\cr}
$$
with
$$
A_1=M_1^2+\delta ^2~~~~~~~~~~~~~~~~~~~~~~~A_2=M_2^2+\delta ^2
\eqno(4c)
$$

With
$$
p_1=xR_1+y'R_2+\pi _1 ~~~~~~~~~~~~~~~~~~p_2=x'R_1+yR_2-\pi _2 
\eqno(5a)
$$
where $\pi _1$ and $\pi _2$ are each transverse to both $R_1$ and $R_2$
and so again are two-dimensional anti-euclidean vectors,
$$
\int d^4p_1\,\delta ^+ (p_1^2)\,\delta ^+ ((P_1-p_1)^2)=
{1\over 4\nu _R}\int dxdy'\,d^2\pi _1\,\delta(y'-\lambda _1(1-x))\,
\delta(\pi _1 ^2+\lambda _1x(1-x))$$$$
\int d^4p_2\,\delta ^+ (p_2^2)\pi ^+ ((P_2-p_2)^2)=
{1\over 4\nu _R}\int dx'dy\,d^2\pi _2\,\delta(x'-\lambda _2(1-y))\,
\delta(\pi _2^2+\lambda _2y(1-y))
\eqno(5b)
$$
Further, 
the energy variable for the $qq$ interaction is
$$
\hat\nu =p_1\cdot p_2= \nu _R\left (xy+\lambda _1\lambda _2(1-x)(1-y)\right )
-\pi _1\cdot\pi _2
\eqno(6a)
$$
and the squared 4-momenta of the virtual quarks are
$$\eqalignno{
\bar p_1^2=(p_1+\Delta)^2&\sim B_1-\beta x +\lambda _1\alpha(1-x)\cr
\bar p_2^2=(p_2-\Delta )^2&\sim B_2-\alpha y +\lambda _2\beta (1-y)
&(6b)\cr}
$$
with
$$
B_1=\delta ^2-2\pi _1\cdot\delta ~~~~~~~~~~~~~~~~~~
B_2=\delta ^2-2\pi _2\cdot\delta 
\eqno(6c)
$$

The $qq$-interaction amplitude $T$ is a function of the energy $\hat\nu$,
the momentum transfer $\hat t$, and the squared 4-momenta $\bar p_1^2$
and $\bar p_2^2$ of the
virtual quarks. According to standard analyticity properties\defref\elop{
R J Eden, P V Landshoff, D I Olive and J C Polkinghorne,
{\sl The analytic S-matrix}, Cambridge University Press (1966)
}, the singularities of $T$ are confined to the upper halves of the complex 
planes of the two variables (6b).  This is conveniently expressed by
the representation
$$
T=\int _0^{\infty}d\kappa _1 d\kappa _2
{\cal T}(\hat\nu,\hat t,\kappa _1,\kappa _2)
\exp i(\kappa _1\bar p_1^2 +\kappa _2\bar p_2^2)
\eqno(7a)
$$
We introduce the $W$ propagators
$$
{\cal P}_1={1\over \bar P_1^2-M_W^2}~~~~~~~~~~~
{\cal P}_2={1\over \bar P_2^2-M_W^2}
\eqno(7b)
$$
insert the expressions (4b) and (6b) into (7) and perform the integrations
over $\alpha$ and $\beta$ by closing each integration contour in the
appropriate half of the complex plane:
$$
\int d\alpha d\beta\; T{\cal P}_1{\cal P}_2=-\int _0^{\infty}d\kappa _1 d\kappa _2
{\cal T}(\hat\nu,\hat t,\kappa _1,\kappa _2)
{(2\pi )^2\over 1-\la _1\la  _2}\theta (f\kappa _1 +G\kappa _2 )\,
\theta (F\kappa  _1+g\kappa _2 )\, \exp i(u_1\kappa _1+u_2\kappa _2 )
\eqno(8a)
$$
where
$$
u_1=B_1+(A_1-M_W^2)f+(A_2-M_W^2)F~~~~~~~~~~
u_2=B_2+(A_1-M_W^2)G+(A_2-M_W^2)g$$$$
f=\la _1\la _2 (1-\L x)/(1-\la_1\la _2)~~~~~~~~~~
g=\la _1\la _2 (1-\L y)/(1-\la_1\la _2)$$$$
F=\la _1 (1-2x)/(1-\la_1\la _2)~~~~~~~~~~
G=\la _2(1-2y)/(1-\la_1\la _2)$$$$
\L={1+\la _1\la _2\over\la _1\la _2}
\eqno(8b)$$

The two $\theta$-functions in (8a) give different limits to the $\kappa _1$
and $\kappa _2$ integrations for different ranges of values of $x$ and 
$y$, corresponding to different signs for $f,g,F,G$. This is shown in
figure 2, where the numbers denote the regions of the ($x,y$)-plane 
for which the integral is non-zero. The region below the dashed curve
is the region $E<0$, where
$$
E=fg-FG
\eqno(8c)
$$
Suppose, as an example, that ${T}(\hat\nu,\hat t,\bar p_1^2,\bar p_2^2)$
were to have the simple factorised form
$$
{T}(\hat\nu,\hat t,\bar p_1^2,\bar p_2^2)=A(\hat\nu,\hat t)
\phi (\bar p_1^2,\mu _1^2)\phi (\bar p_2^2,\mu _2^2)$$$$
\phi (p^2,\mu ^2)={1\over p^2-\mu ^2}
\eqno(9)
$$
Then the integral (8a) is $(2\pi )^2\,A(\hat\nu,\hat t)/(1-\la _1\la _2)$ 
times the following:

\topinsert
\centerline{\hbox{\epsfxsize=55truemm\epsfbox{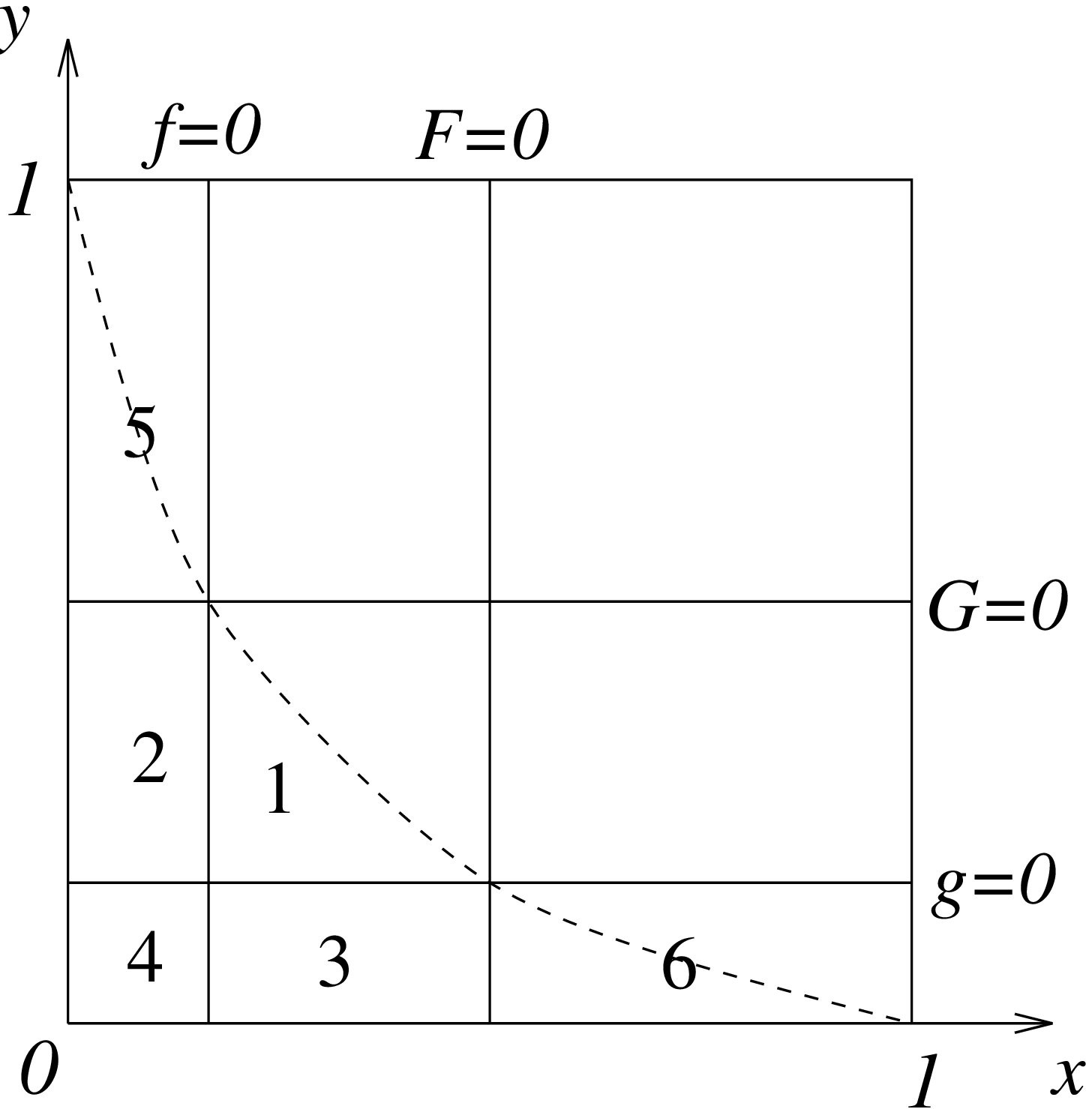}}}
Figure 2: Regions of the ($x,y$)-plane for which the integral (8a) is
nonzero. The curve is $E=0$, with $E$ given in (8c)
\endinsert
{\leftskip 10truemm
{\bf Region 1}: $F,G>0,~~~f,g<0$\h
$$
{E\theta (-E)\over FG}{1\over D_1D_2}
$$
{\bf Region 2}: $F,G>0,~~~f>0,~g<0$~~~~~~~~(so that $E<0$)\h
$$
-{1\over U_1D_2}
$$
{\bf Region 3}: $F,G>0,~~~f<0,~g>0$~~~~~~~~(so that $E<0$)\h
$$
-{1\over D_1U_2}
$$
{\bf Region 4}: $F,G>0,~~~f,g>0$~~~~~~~~(so that $E<0$)\h
$$
-{1\over U_1U_2}
$$
{\bf Region 5}: $F>0,~G<0~~~f>0,~g<0$\h
$$
-\theta (E){C\over U_1D_1}-\theta (-E){1\over U_1D_2}
$$
{\bf Region 6}: $F<0,~G>0~~~f<0,~g>0$\h
$$
-\theta (E){c\over U_2D_2}-\theta (-E){1\over D_1U_2}
\eqno(10a)
$$\par}
where
$$
U_1=u_1-\mu _1^2~~~~~~~~~~~~~~~~~~~~~U_2=u_2-\mu _2^2$$$$
D_1=U_1+CU_2$$$$
D_2=cU_1+U_2$$$$
C=-f/G~~~~~~~~ c=-g/F
\eqno(10b)$$
This complexity is not unexpected. In coordinate space there are many
different contributions, corresponding to the possible different time-%
orderings of the vertices in figure 1b.  Although we cannot identify a direct
correspondence between the different time-orderings and the various cases
in (10), we speculate that there is a connection between the two.

For both singlet and octet exchange we
perform the integrations over the two components of $\delta$ numerically.
In each case we find that
there is an important contribution from values of $x$ and $y$ that lie
near the curve $E=0$. When $E=0$,
$$
GD_1=GB_1-fB_2~~~~~~~~~~~~~FD_2=FB_2-gB_1
$$
and so the dependence of $D_1$ and $D_2$ on $M_W$ disappears. So, on the
curve $E=0$ in region
1, the amplitude of figure 1b no longer depends on the $W$ mass: presumably
this corresponds to the $W$'s decaying very quickly, before they can
propagate, so that the quarks are still close together and their interaction
is therefore enhanced\ref{\sk}.
\bigskip
{\bf Soft-pomeron exchange}

In the case of colour-singlet exchange, we find that the amplitude is
so strongly peaked near to $E=0$ that we do not need to consider interference
between the different ways of attaching the exchange to the quark
lines. We calculate $R$, defined in (1) and appropriately summed over
the different ways of attaching the exchanges to the quarks, and
integrate $|1+R|^2$ over the angle between $\pi$ and $\pi '$.
So we include the square of figure 1b, and also the interference with the
Born term of figure 1a. In this way we determine the ratio 
$$
{\cal R}(M_1,M_2,x,y)=
{d^4\sigma ^{\hbox{\sevenrm CORRECTION}}\over dM_1dM_2dxdy}\Big / 
{d^4\sigma ^{\hbox{\sevenrm BORN}}\over dM_1dM_2dxdy} 
\eqno(11)
$$
where $x$ and $y$ are defined in (5a).

We now introduce the specific soft-pomeron-exchange form for the final-%
state interaction. Then\ref{\zuoz}
$$
A(\hat\nu,\hat t)=2\beta _0^2\,(2\hat\nu )^{\alpha (\hat t)}$$$$
\alpha (\hat t)=1+\epsilon _0+\alpha '\hat t$$$$
\beta _0=2\hbox{ GeV}^{-1} ~~~~~~~  \epsilon _0=0.08 ~~~~~~~~\alpha '=0.25
\hbox{ GeV}^{-2}
\eqno(12a)
$$
We need a form factor for the coupling of the pomeron to the off-shell
quarks. The best available  choice is\defref\condensate{
A Donnachie and P V Landshoff, Nuclear Physics B311 (1989) 509
}
to use (9) and replace the function $\phi (p^2,\mu ^2)$ with
$$
\phi (p^2,0)-\phi (p^2,\mu _0  ^2)
\eqno(12b)
$$
with $\mu _0\approx 1\hbox{GeV}$.

The result of our numerical computation of ${\cal R}$, defined in (11), is
that it is extremely small over almost the whole $(x,y)$ plane, at
the one per mil level or less. This is to be attributed to colour
transparency\ref{\bertsch}, which in our calculation manifests itself
as a strong cancellation between the two terms in (12b). 
\bigskip
\pageinsert
\centerline{{\epsfxsize=95truemm\epsfbox{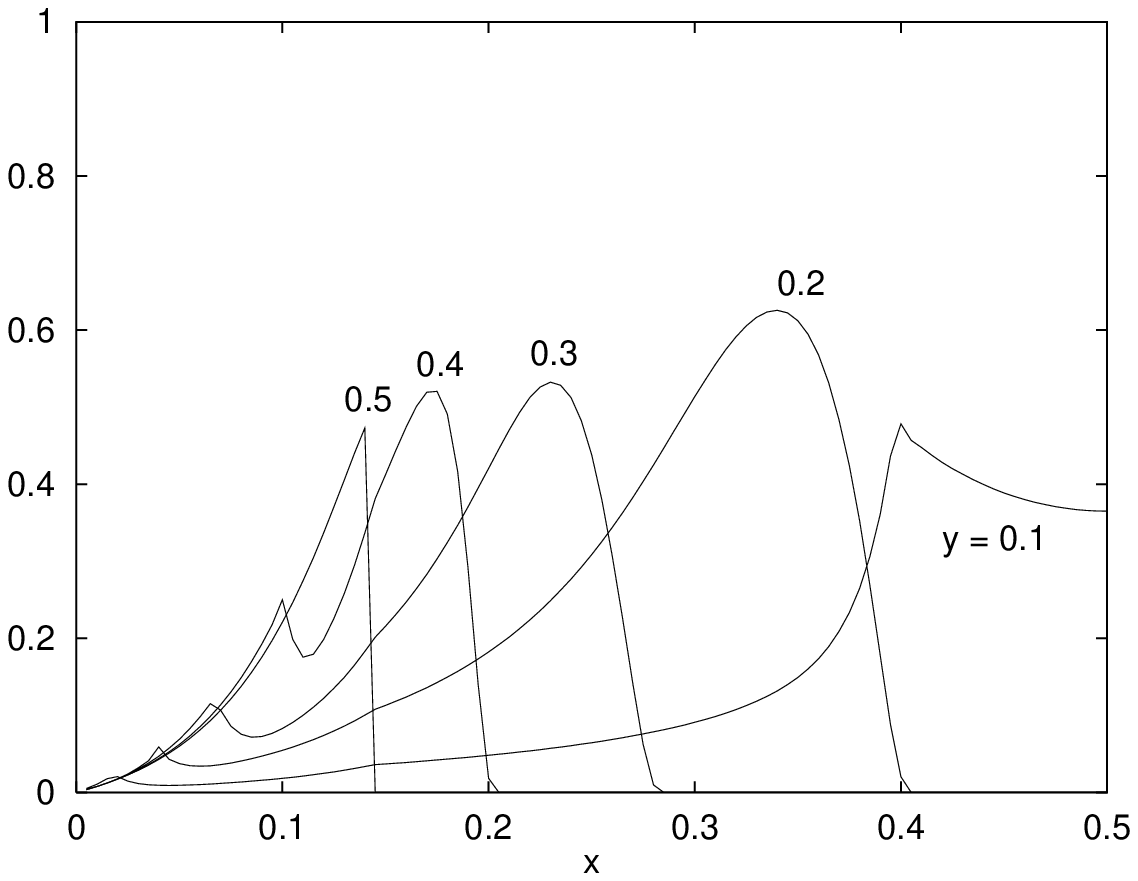}}}
\centerline{(a)}

\centerline{{\epsfxsize=95truemm\epsfbox{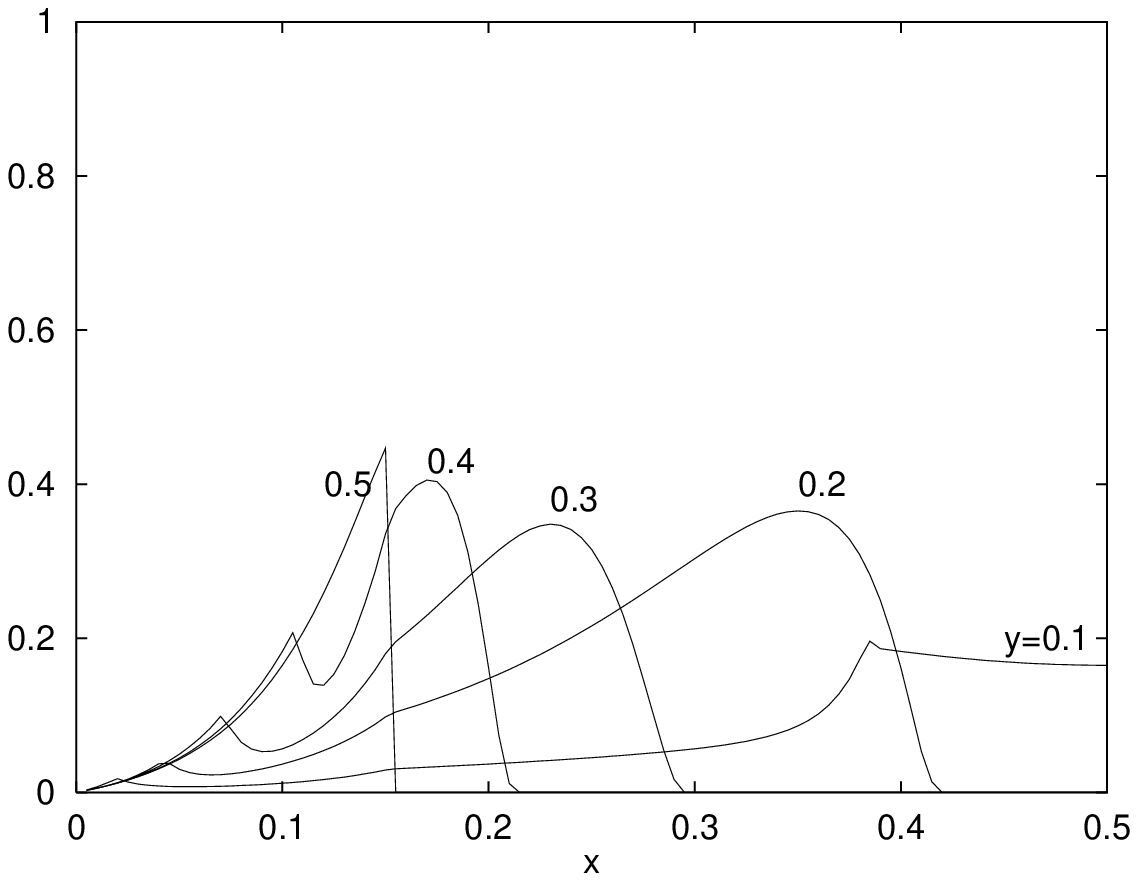}}}
\centerline{(b)}

\centerline{{\epsfxsize=95truemm\epsfbox{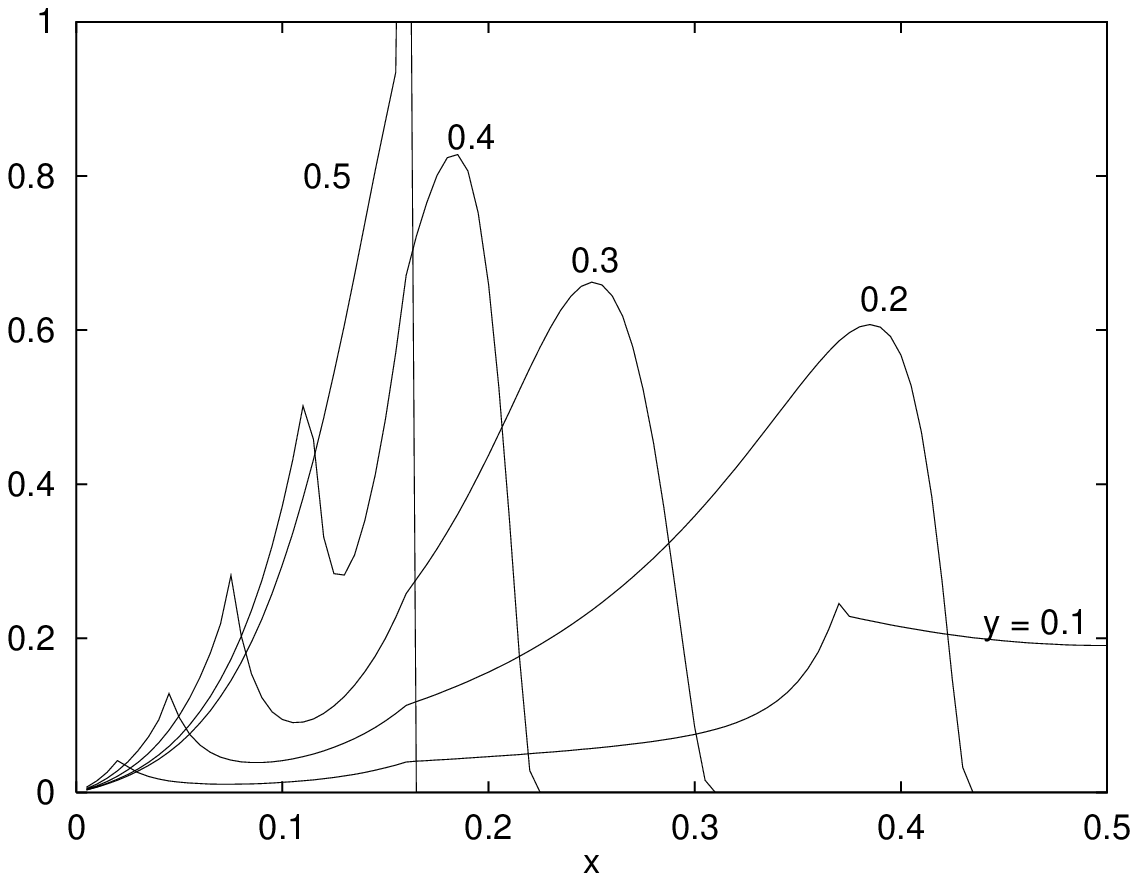}}}
\centerline{(c)}
Figure 3: The ratio $\cal R$, defined in (11), at $\surd s=175$ GeV
and $M_1=M$, for (a) $M_2=M$, (b)~$M_2=M-\Gamma$, and (c)
$M_2=M+\Gamma$.
\endinsert

{\bf Octet exchange}
\def\q{{\bf q}}

According to Cornwall's calculation\ref{\cornwall}, colour-octet exchange
between quarks 
can be well approximated by 
$$
A(\hat\nu ,\hat t)=16\pi \hat\nu\,\alpha _s(-\hat t)D(-\hat t)
\eqno(13a)
$$ 
with
$$
D^{-1}(\q ^2) = \q ^2 + m^2(\q ^2)$$$$
\alpha_{s}(\q ^2) = {{12\pi}\over{(33-2N_{f})\log\Bigl[{{\q ^2+4m^2(\q ^2)}\over
{\Lambda^2}}\Bigr]}}\eqno(13b)
$$
where the running gluon mass is given by
$$
m^2(\q ^2) = m_0^2\Biggl[{{\log{{\q ^2+4m_0^2}\over{\Lambda^2}}}\over{\log{{4m_0^2}
\over{\Lambda^2}}}}\Biggr]^{-12/11}
\eqno(13c)
$$
The fixed mass $m_0^2$ can be determined\defref\halzen{
A Donnachie and P V Landshoff, Nuclear Physics B311 (1988/89) 509\h
M B Gay Ducati, F Halzen and A A Natale, Physical Review D48 (1993) 2324
} from the condition that the
simple exchange of a pair of gluons between quarks is the zeroth-order
approximation to soft pomeron exchange at $t = 0$. This requires that
the integral
$$
\beta_0^2 = {{4}\over{9}}\int d^2\q \,[\alpha_s(\q ^2)D(\q ^2)]^2
\eqno(14)
$$
be about 4 GeV$^{-2}$. With a choice of $\Lambda = 200$ MeV this gives
$m_0 = 340$ MeV.

While, strictly speaking, Cornwall's analysis applies only to Euclidean
$\q$, the form of (13) suggests that it effectively gives the gluon a
mass which at low momentum values is close to $m_0$. Thus, while there is
uncertainty about how to calculate gluon emission, it seems that the
confinement effects that Cornwall's calculation reveals do not allow
the emission of soft gluons. So, while the effect of gluon emission
tends to cancel that of gluon exchange, in the case of soft exchange
there is no such cancellation. For this reason, we calculate the
contribution from exchange for which $\q ^2<m_0^2$, and suggest that
this provides a lower bound to the correction to the cross section
from colour-octet exchange and emission. While we cannot be certain
that this is a meaningful approach, we observe that the calculation is
a very different one from that of the corresponding correction to the
$e^+e^-$ total cross section. In the $WW$ calculation there are
$W$ propagators in addition to quark propagators, which damp the contribution
to the integration (3b) from large values of $\alpha$ and $\beta$, so
that the momentum transfer $\Delta ^2$ carried by the exchanged gluon is
confined to spacelike values, as is seen in (4a). In the $e^+e^-$ case
there is no such damping, and hence there are additional contributions from
timelike $\Delta ^2$, which we cannot calculate because the Cornwall form
(13) of the gluon-exchnage amplitude is valid only for spacelike $\Delta ^2$.

We calculate the sum of the amplitudes of figure 1b with the soft-gluon
exchange attached to the quarks in all of the four possible 
ways\footnote{$^*$}{We do not consider diagrams where the gluon is
exchanged simply between a pair of quarks associated with the same
$W$, since this is just the same as the familiar correction to
$R_{e^+e^-}$ and so is known to be small.}.
We square this amplitude: colour considerations forbid interference with
the Born term of figure 1, and of course in the squared amplitude there
appears a colour factor 2/9 because the two gluons together must form
a singlet configuration. 
We use the form (13a) for $A$ in (9). We simply set $\mu _1=\mu _2=0$,
because the nonperturbative
quark propagator corresponding to the Cornwall gluon propagator
is not available.
The output for the ratio ${\cal R}$, defined in (11), is symmetric under
$x\to (1-x)$ or $y\to (1-y)$ (or both). Figure 3 shows the contribution to
${\cal R}$ from the exchange of nonperturbative 
gluons with $\q ^2 <0.1$ GeV$^2$. 
The energy is $\surd s=$ 175 GeV and the plots  are for $M_1=M$, 
with $M_2=M$ and 
$M\pm\Gamma$. The rather violent dependence on $x$ and $y$ is striking. 

\topinsert
\centerline{{\epsfxsize=110truemm\epsfbox{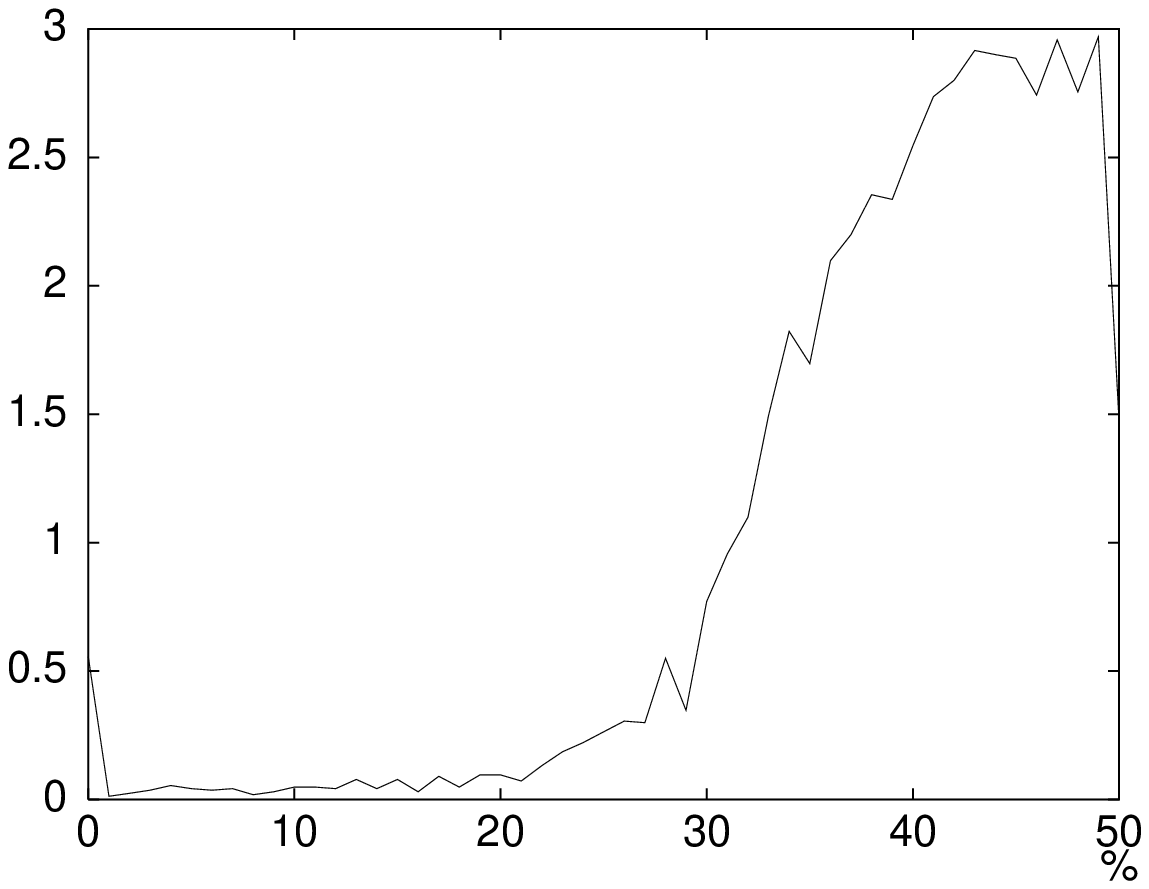}}}
Figure 4: Probability distribution for fractional energy of jets in the
Born approximation
\endinsert
\def\R{{\cal R}}
The interaction has a significant effect on the energy distribution  of the 
jets, though the overall effect on the integrated cross section is not so large.
To convert the plots of figure 3 to energy distributions, note that the
energy of the jet $p_1$ is $\half X\sqrt{s}$, with
$$
X={x(1-2\la _1+\la _1\la _2)+\la _1-\la _1\la _2 \over 1-\la _1\la _2}
\eqno(15)
$$
There are similar equations for the fractional energies of the other three
jets. In the Born approximation, the probability distributions of the two
pairs of jets are uncorrelated, and each is symmetric under $X\to (1-X)$.
Figure 4 shows the output, slightly smoothed, of a Monte Carlo 
calculation\defref\monte{
D Eatough and T Wyatt, private communication} of the 
probability distribution $P(X)$ at energy $\sqrt{s}=172$ GeV. 
We use this distribution
to weight the output $\R $ defined in (11) and we then average it over $x$
and $y$. The result is shown in figure 5.
\bigskip
\centerline{\epsfxsize=80truemm\epsfbox{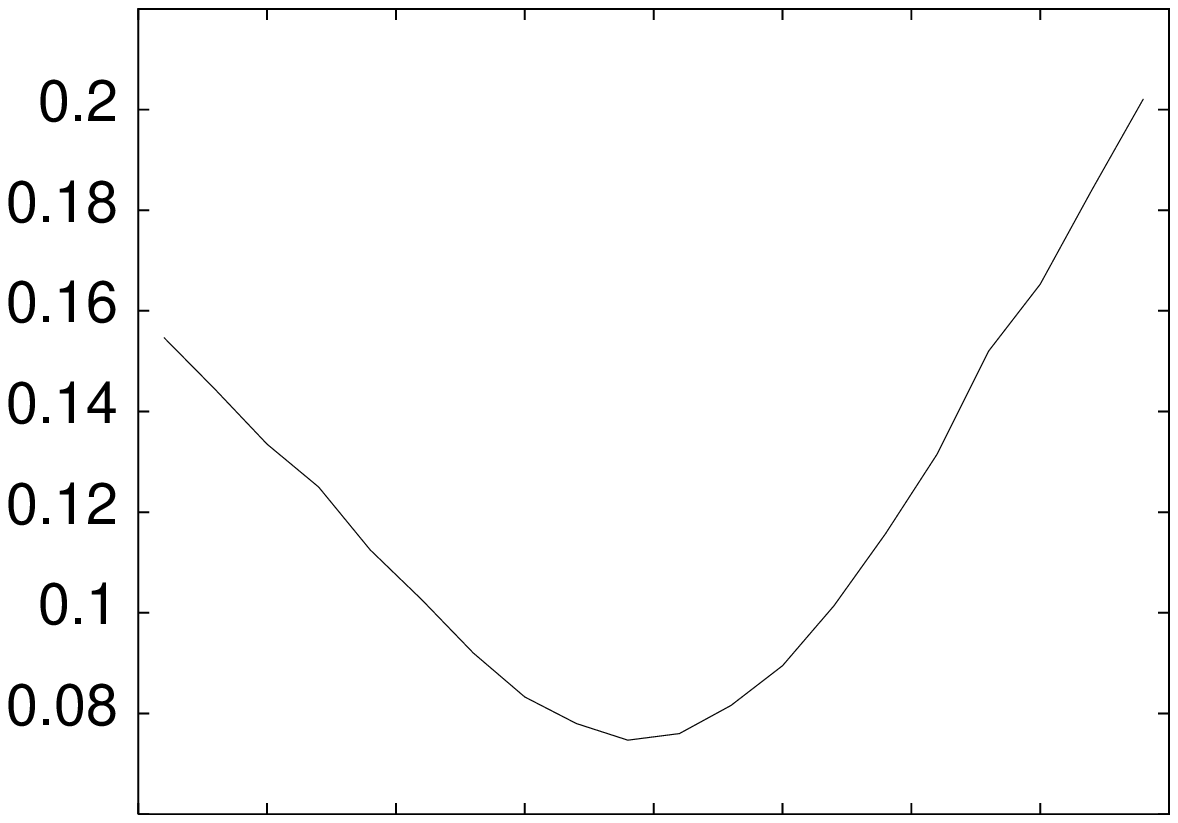}}
\vskip -2mm
\centerline{\hskip 7mm $M-\G$\hskip 23mm $M$ \hskip 23mm $M+\G$}
Figure 5: $\langle{\cal R}\rangle$ for $\surd s=$ 175 GeV, $M_1=M$,
plotted against $M_2$.
\bigskip
{\bf Conclusions}

We have found that colour transparency causes
colour-singlet exchange to have very  little effect on
the cross section and spectra for the process $e^+e^-\to WW\to q\bar q q\bar q$.
We have calculated this by modelling colour-singlet in terms of a
phenomenological soft-pomeron exchange, in which the soft pomeron couples
to quarks with a form factor that goes to 0 when the quarks go far off shell.
We have shown elsewhere\ref{\condensate} that when soft-pomeron exchange is
modelled in terms of the exchange of a pair of nonperturbative gluons, this
form factor arises from taking account of the various attachments of the
gluons to the separate quarks of a colour-singlet system, that is it arises
from colour-transparency effects.
 
For colour-octet exchange the situation is very different.  We calculate 
the contribution to this from soft-nonperturbative-gluon exchange, which
we argue to give a lower bound to the total colour-octet exchange. Figure 
5 shows that this will make a small but noticeable change in the total rate
for the $q\bar q q\bar q$ final state. In some regions of phase space the
local effect can be significant, as can be seen from figure 3, and the
energy distribution of jets will differ from that of the Born term alone. 
However there will be little impact on the mass determination due to the near
symmetry of figure 5, although some increase in the width can be anticipated.

\bigskip\bigskip
{\sl We are grateful to David Eatough and Terry Wyatt of the OPAL 
Collaboration for providing us with their four-fermion Monte Carlo
output.
 
This research was
supported in part by the UK Particle Physics and
Astronomy Research Council and by the EC Programme
``Training and Mobility of Researchers", Network
``Hadronic Physics with High Energy Electromagnetic Probes",
contract} ERB FMRX-CT96-0008.

\vfill\eject
\medskip\immediate\closeout\rfile\writestoppt
\baselineskip=14pt{{\bf References}}\par{\frenchspacing%
\parindent=20pt\escapechar=` \input refs.tmp\bigskip}\nonfrenchspacing
\bye